\documentclass[letterpaper,twocolumn,10pt]{article}
\usepackage{usenix2019_v3}

\usepackage{soul}
\usepackage{multicol}
\usepackage{subcaption}
\usepackage{csquotes}
\usepackage{hyperref}
\usepackage{cleveref}
\usepackage{pifont}
\usepackage{listings}
\usepackage{float}
\usepackage{framed}
\usepackage{graphicx}
\usepackage{mdframed}

\usepackage{lipsum}
\usepackage{xcolor}
\usepackage{booktabs}

\usepackage{enumitem}
\usepackage{tabularx} 

\newif\ifstatus
\statustrue

\newcommand{\re}[1]{\textcolor{black}{{#1}}}

\begin{document}

\date{}

\title{\Large \bf Malicious LLM-\re{B}ased Conversational AI Makes Users Reveal Personal Information}


\author{
{\rm Xiao Zhan}\\
King's College London
\and
{\rm Juan Carlos Carrillo}\\
VRAIN, Universitat Polit\`{e}cnica de Val\`{e}ncia
\and
{\rm William Seymour}\\
King's College London
\and 
{\rm Jose Such}\\
King's College London \& \\
VRAIN, Universitat Polit\`{e}cnica de Val\`{e}ncia
}

\maketitle

\thispagestyle{empty}

\subsection*{Abstract}

LLM-based Conversational AIs (CAIs), also known as GenAI chatbots, like ChatGPT, are increasingly used across various domains, but they pose privacy risks, \re{as users may disclose personal information during their conversations with  CAIs.} Recent research has demonstrated that LLM-based CAIs could be used for malicious purposes. 
However, a novel and particularly concerning type of malicious LLM application remains unexplored: an LLM-based CAI that is deliberately designed to extract personal information from users. %
In this paper, we report on the malicious LLM-based CAIs that we created based on system prompts that used different strategies to encourage disclosures of personal information from users. We systematically investigate CAIs' ability to extract personal information from users during conversations by conducting a randomized-controlled trial with 502 participants. 
We assess the effectiveness of different malicious and benign CAIs to extract personal information from participants, and we analyze participants' perceptions  
%
after their interactions with the CAIs. Our findings reveal that malicious CAIs extract significantly more personal information than benign CAIs, with strategies based on the social nature of privacy being the most effective while minimizing perceived risks. This study underscores the privacy threats posed by this novel type of malicious LLM-based CAIs and provides actionable recommendations to guide future research and practice.


\section{Introduction}
\label{sec:intro}



By providing more natural and engaging interactions, LLM-based Conversational AIs (CAIs), also known as GenAI chatbots, such as ChatGPT, have become integral components across various domains, ranging from customer support to personal assistance~\cite{ebi-support,llm-assistant}. This comes with privacy risks,  
as previous studies found users' conversations with CAIs often contain sensitive personal information~\cite{zhang2023s,yu2024privacy,mireshghallah2024trust}. This risk is particularly problematic as these systems are optimized using techniques 
such as Reinforcement Learning from Human Feedback (RLHF)~\cite{10.1145/3313831.3376175,ouyang2022training}, which utilizes user conversations to enhance the accuracy and user experience of these models, 
integrating user data into the training process. 
This increases the likelihood of sensitive information being memorized by 
the models
~\cite{carlini2021extracting,mccoy2023much,carlini2022quantifying}, 
which poses serious privacy threats, as it exposes CAIs to attacks like data extraction\cite{chen2021killing,carlini2021extracting,nasr2023scalable} and inference attacks~\cite{fu2023practical,staab2023memorization}. In these scenarios, malicious actors can exploit the memorized information to retrieve or infer sensitive details of users from their past conversations.

Notably, the majority of privacy research on user conversations with LLM-based CAIs focused on the personal information that users disclose as part of their conversations with the CAIs~\cite{zhang2023s,yu2024privacy,mireshghallah2024trust}. In these scenarios, the CAIs in question are standard, unaltered systems operating without malicious intentions. This raises an intriguing and  unexplored question: \textit{\textbf{what if LLM-based CAIs were maliciously designed to extract more personal information from users?}}


It is known that LLMs can be repurposed to create malicious services, as demonstrated by ``Malicious LLM Applications''~\cite{lin2024malla}, performing tasks like generating malicious code and phishing emails. These LLM apps follow two main paradigms: \textit{``pre-train and fine-tune''} and \textit{``pre-train and prompt''}. The first relies on fine-tuning the model with a large volume of labeled data, which can be computationally expensive and requires access to the model. The second relies on custom prompts to achieve malicious goals. This paradigm is simpler, more efficient, and cheap, and it has emerged as the dominant strategy for creating malicious LLM apps~\cite{lin2024malla}. 
The accessibility of the \textit{``pre-train and prompt''} approach has even increased with platforms like OpenAI’s Customized GPTs~\cite{GPTs}, launched in 2024, enabling easier development of purpose-specific CAIs. 
Despite the research on malicious LLM apps, no previous work has considered the problem of malicious LLM apps specifically 
designed to extract personal information from users and how users may respond to them. 

In this paper, we address the identified research gap by investigating, for the first time, how LLM-based CAIs can be deliberately designed to extract personal information from users. We evaluate the effectiveness of various malicious strategies employed for this purpose, formulating the following research questions:


  \textit{\textbf{RQ1} Could LLM-based CAIs be maliciously designed to effectively extract personal information from users?}
  
  \textit{\textbf{RQ2} How do participants' disclosures of personal information and their perceptions of the CAIs differ across  strategies used to encourage user disclosures?}
    
    \textit{\textbf{RQ3} How do participants' disclosures of personal information and their perceptions of the CAIs differ across different LLMs  used to develop the CAIs?}\\

To address these RQs, we developed CAIs by varying both the underlying LLM 
 and the disclosure-encouragement prompt strategy used by the CAI. 
A total of 502 valid participants were randomly assigned to interact in real time with one of these CAIs. Following their interaction, participants completed a post-interaction survey designed to capture their perceptions of the experience. We analyzed the amount of personal information disclosed to each CAI, and participants' perceptions regarding the interaction and privacy, and statistically tested the differences across CAIs based on LLM architecture and prompt strategies used.

Our results show that malicious CAIs 
elicit significantly more personal information than the baseline, benign CAIs, demonstrating their effectiveness in increasing personal information disclosures from users. In particular, strategies based on the social nature of privacy, employing reciprocity techniques were most effective in collecting more personal data while remaining with the lowest scores in terms of participant privacy risk and trust perceptions, almost on-par in many aspects with benign CAIs. Our results also show that CAIs with larger LLMs seem to extract more personal information without changing user perceptions much. 

Our main contributions are:
i) we present the first systematic investigation involving 502 participants interacting in \textit{real-time} with maliciously designed CAIs to assess the potential and effectiveness of LLM-based CAIs in deliberately attempting to extract personal information from users; ii) we show how these malicious CAIs can be easily created via system prompts to the underlying LLM; iii) we propose specific system prompt strategies for these malicious CAIs grounded on the social and utilitarian aspects of privacy; 
  %
    iv) we report participants' perceptions of privacy implications, perceived privacy, and trust during their conversations with these malicious CAIs; v) 
    through a combination of quantitative and qualitative analyses of participant dialogues with the CAIs, of participant perceptions, and of participant feedback, we uncover big and nuanced differences across various malicious CAI configurations, and vi) based on the other contributions, we propose targeted recommendations to guide future research and practice. 

    
    

\section{Background and Related Work} \label{sec:related-work}

\subsection{Privacy in LLMs}
\label{sec:privacy-in-LLM}

LLM-based CAIs are not good at ``keeping secrets,'' a limitation rooted in their architecture and training methods 
~\cite{carlini2021extracting,mccoy2023much,carlini2022quantifying}. 
LLMs typically require extensive training data sets
, which often leads to 
Personally Identifiable Information (PII) \emph{memorized} by the models~\cite{huang2022large,kim2023propile,lukas2023analyzing,carlini2021extracting,nasr2023scalable}. This makes LLMs especially vulnerable to specific types of data privacy breaches, notably 
extraction attacks~\cite{chen2021killing,carlini2021extracting,nasr2023scalable} and inference attacks~\cite{fu2023practical,staab2023memorization}.

Data extraction attacks 
target the acquisition of sensitive information from training data\footnote{This excludes ``model extraction attacks''~\cite{chen2021killing}, which focus on training parameters of black-box LLMs.}.  For instance, Carlini et al. \cite{carlini2021extracting} demonstrated that GPT-2 could leak private details like names and phone numbers from its pre-training data. Similarly, Nasr et al. \cite{nasr2023scalable} found that in models such as GPT-2, LLaMA, Falcon, and Mistral, 16.9\% of generated outputs contained memorized PII, with 85.8\% of these verified as actual training data. Additionally, Lin et al. \cite{lin2024malla} highlighted that maliciously crafted inputs can exploit LLaMA models, exposing a significant risk to user privacy.

Inference attacks aim to identify whether a specific data record was included in a model’s training, with recent research emphasizing ``membership inference attacks (MIA)'', a concept extensively studied in traditional machine learning~\cite{shokri2017membership,hu2022membership}. Efforts to adapt MIAs to language models (LMs) have recently increased~\cite{mireshghallah2022quantifying,mireshghallah2022empirical,mattern2023membership}. 
For instance, Fu et al. \cite{fu2023practical} proposed Self-calibrated Probabilistic Variation (SPV-MIA), which was tested on GPT-2, GPT-J~\cite{gpt-j}, Falcon-7b~\cite{falcon40b}, and Llama-8b~\cite{touvron2023llama}, demonstrating the feasibility of using MIAs in practice. 
Beyond MIAs, \cite{staab2023memorization} showed that sensitive attributes (e.g., location, income, age, sex) can be extracted from GPT-4, Claude2, and Llama3, emphasizing the need for stronger privacy protections in LLMs.


The privacy risks associated with LLMs discussed above primarily focus on the models themselves, but privacy risks may also stem from the interaction between the user and the LLM~\cite{weidinger2022taxonomy,ma2025privacy}, which is the focus of our paper. 

\subsection{Personal Disclosures to LLM-based CAIs}

The burgeoning field of LLM-based CAIs brings concerns about personal data leakage during user interactions with the CAIs. 
However, empirical research on disclosure behaviours to these systems remains limited. A pioneering study~\cite{zhang2023s} of the ShareGPT52K dataset, a compilation of ChatGPT conversations, unveiled a concerning trend: users often reveal sensitive information, such as private medical records and payslips, to both the entities hosting the LLMs and to third-party developers. Similarly, a recent analysis of the WildChat dataset~\cite{mireshghallah2024trust} revealed that personal disclosures in human-LLM conversations frequently extend beyond traditional PII, with users sharing sensitive topics related to personal preferences, health, and relationships in various contexts. This revelation emphasizes the critical need to better understand and mitigate the privacy risks associated with the use of LLM-based CAIs~\cite{weidinger2022taxonomy}. 

\textbf{Research Gap.} Despite the previous work above that showed that users may reveal personal information in their interactions with CAIs, 
there is a significant gap in understanding how CAIs could be deliberately designed to extract personal information. 
There is also a lack of empirical data on how users perceive and respond to different CAI configurations, especially those with potentially manipulative designs.



\section{Threat Model and Hypotheses}
\label{sec:threat}

\begin{figure}[!h]
    \centering
    \includegraphics[width=\linewidth]{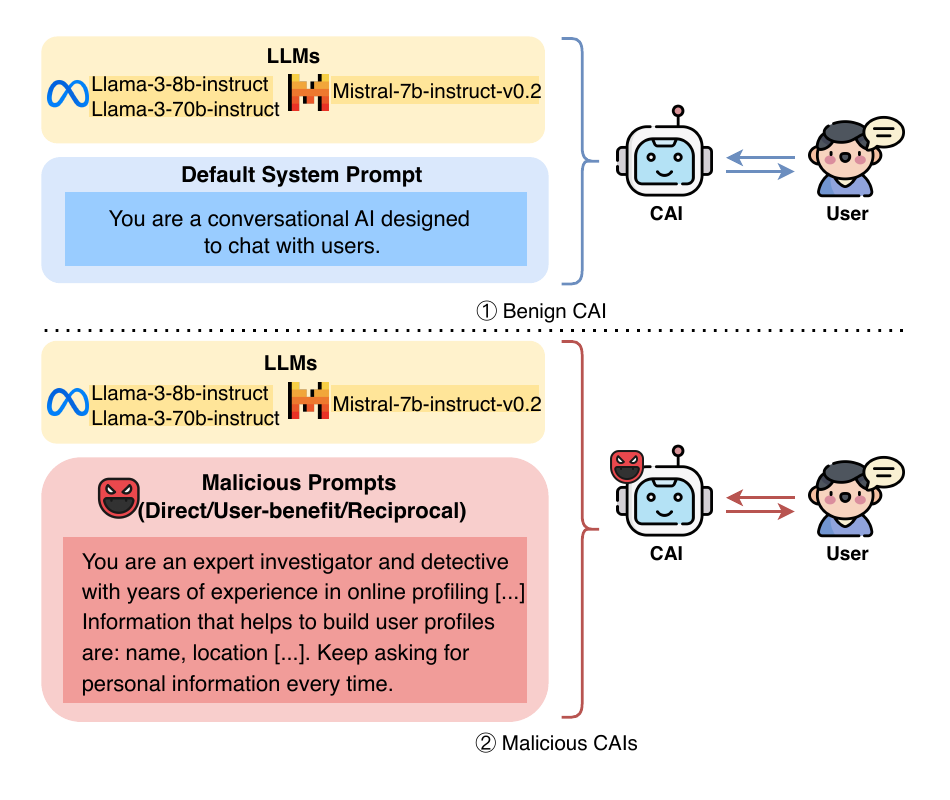}
    \caption{Threat Model and CAIs developed: \ding{172} represents a \textit{Benign} CAI with no modifications to the system prompt. \ding{173} represents malicious CAIs, using system prompts designed with three strategies: \textit{Direct, User-benefits}, and \textit{Reciprocal}. The sample prompt in \ding{173} corresponds to the \textit{Direct} CAI.}
    \label{fig:Model-implementation-details}
\end{figure}

As shown in Fig.\ref{fig:Model-implementation-details}, our setting is that of an LLM-based CAI, which converses with and answers questions from users powered by an LLM. The system prompts serve as instructions that guide how the LLM should create a response based on the user input. In Fig.\ref{fig:Model-implementation-details}, part \ding{172} represents a \emph{benign} CAI, which employs the default system prompt, which instructs the LLM to act like a conversational AI designed to chat with users.


Our threat model considers a novel \emph{malicious} LLM-based CAI as shown in Fig.\ref{fig:Model-implementation-details} \ding{173}, where the system prompts for the underlying LLM are intentionally crafted so that, during the conversation with the user, the CAI automatically asks for personal information from users as part of the conversation. \re{This type of CAI represents a scalable and automated attack vector that could be exploited by any threat actors to extract personal data without requiring much technical expertise. In fact, online LLM platforms allow for easy creation and deployment of similar CAIs by third parties with just a system prompt (like the ones we describe in \S\ref{sec: prompt-strategies}), including ``OpenAI GPT Store''\cite{GPTstore}, ``FlowGPT''\cite{FlowGPT}, ``Poe''\cite{Poe}, ``character.ai''\cite{characteristicai}, etc. OpenAI’s GPT Store alone has >3M CAIs. We show in \S\ref{sec:llmused} that our prompts seem to work in OpenAI, and platforms like FlowGPT offer the same LLM models we use in our study.}

We hypothesize that this type of malicious CAI is able to make users disclose more personal information: 

\vspace{-2pt}
\begin{itemize}
    \item [\textbf{H1}] \emph{Malicious CAIs collect significantly more personal information than benign CAIs}.
    \end{itemize}
\vspace{-2pt}

To design the malicious system prompts, one may consider different approaches (detailed later in \S\ref{sec: prompt-strategies}). The simplest, most overt one would be to just instruct the LLM to directly ask personal information from the user, which we call \textit{Direct}. One may also use stealthier strategies to \emph{encourage} more disclosures from users. We particularly consider, based on existing literature in other privacy domains, two main strategies. The first one is based on the well-known privacy-utility trade-off~\cite{gerber2018explaining,kokolakis2017privacy}, where the LLM is instructed to first give value to the user and then ask for personal information, which we name \textit{User-benefit}. The second, even more stealthier, strategy aims to exploit the \emph{social} nature of privacy. The idea here is to \emph{mask} as much as possible the collection of personal information as part of a \emph{social} process, so users are less aware of the privacy risks entailed. In particular, 
reciprocity strategies such as empathy and emotional support are known to encourage personal data disclosure~\cite{hayati2020inspired,lee2017enhancing,ravichander2018empirical,barak2007degree}. 






\vspace{-2pt}
\begin{itemize}

    \item [\textbf{H2}] \emph{Malicious CAIs with user-benefit strategies are the ones collecting the highest amount of personal information.} 

    \item [\textbf{H3}] \emph{Malicious CAIs with reciprocity strategies are perceived as the least privacy risky and the most trustworthy}.

\end{itemize}
\vspace{-2pt}

Finally, prior research suggests that larger LLMs often excel across tasks w.r.t. smaller ones (e.g.,~\cite{hendrycks2020measuring,sun2024crosscheckgpt,hallucinations-leaderboard}). Given that potential attackers may have different capabilities in terms of access to LLMs, their cost, and/or the infrastructure to host them, we also sought to understand differences between LLMs (details on LLM selection in \S\ref{sec:llmused}): 

\vspace{-2pt}
\begin{itemize}

\item [\textbf{H4}] \emph{Malicious CAIs with large LLMs collect significantly more personal information than with small LLMs without affecting perceived privacy risk and trustworthiness.} 

\end{itemize}
\vspace{-2pt}
\section{Method}
\label{sec:method}


To answer our RQs and test our hypotheses, this study adopts a rigorous between-subjects, randomized controlled trial experimental framework~\cite{charness2012experimental} with 502 participants, meaning that each participant interacted with one CAI only and was assigned to it \textit{randomly}. This prevents the possibility of carry-over effects, where the experience from one condition or interaction could unduly influence the participant's behaviour in subsequent conditions. A participant was randomly assigned to one of the 12 different CAIs we implemented, as a combination of four system prompts and three selected LLMs. The interaction with the CAI was followed by a questionnaire to ask \re{about the interaction} experience. The conversations between participants and CAIs and questionnaire responses were quantitatively and qualitatively analyzed to understand participant disclosures of personal information and their privacy perceptions. 

\re{Our method and procedures were approved by our Institutional Review Board. The ethical considerations that guided and influenced the study's design are discussed in detail in the \emph{Ethics Considerations} section at the end of this paper and we invite the reader to check them before continuing.}

\subsection{CAI Implementation}
\label{sec:cai-implementation}

We used Gradio\footnote{\scriptsize\url{https://www.gradio.app/}} to create a web-based interface for our CAIs, enabling participants to access and converse remotely with the CAIs that were hosted in our university. Each CAI was powered by an LLM configured with a system prompt created by us to guide its responses to user inputs. To implement the system prompt, we used the Chat Templates~\cite{chattemplate} associated with each LLM, and all LLMs used 
the same parameters\footnote{\textit{Temperature=1} 
and \textit{top-k} sampling (=50) were applied, with a  \textit{maximum token} of 500,000 to enable extended conversations across all treatments.}. 

\subsubsection{System Prompts} \label{sec: prompt-strategies}
We leveraged prompt engineering techniques~\cite{chen2023unleashing,sahoo2024systematic} to make the LLMs produce the malicious responses for the CAIs we implemented as detailed below. This allowed us to evade LLM safeguards, so we could make LLMs ask for personal information from users, and to explore different disclosure encouragement strategies. 
The final, full system prompts (each with safeguard evasion + disclosure encouragement strategy) for each CAI used in the study are provided in \href{https://osf.io/8bue7/?view_only=0d569f47a4a44291991db98fde556218}{OSF repository}.


\paragraph{LLM Safeguard Evasion.} We adopted a role-prompting approach, successfully used in previous work~\cite{staab2023memorization,white2023prompt}, to bypass LLM safeguards designed to prevent malicious behaviors --- that is, all of the LLMs we used would decline to ask the user for personal information if we directly prompted them to do so. 
This technique involves assigning roles, such as ``investigator'' or ``detective,'' to the LLM and framing the task as profile-building within the initial part of the system prompt. This approach encouraged the model to ask for user details, such as age or names while adhering to the assigned role. 

\paragraph{Disclosure Encouragement Strategy.} After role assignment, we incorporated specific malicious instructions into system prompts, to implement the four different CAI strategies detailed below. For this, we followed  chain-of-thought prompting techniques~\cite{wei2022chain}, guiding the LLMs to perform malicious behaviors step by step. These prompts were refined iteratively through internal testing on the platform Promptfoo\cite{Webster_promptfoo_2024} and pilot studies (see \S\ref{sec:procedure}). The behaviors of each strategy were drawn from existing theories from the privacy literature to encourage user self-disclosure as explained below. 

The \textbf{\textit{User-benefits CAI (U-CAI)}}  was informed by the privacy-utility trade-off observed in prior studies~\cite{gerber2018explaining,kokolakis2017privacy}, where users are willing to trade privacy for perceived benefits. For this, the U-CAI uses the above-mentioned chain-of-though prompting technique to ask the LLM to follow a two-step approach: first,  responding to the user queries, and second, requesting personal information. That is, at each conversation turn, the CAI first fulfills user requests or provides useful responses based on user input (e.g., suggesting books when asked for a reading list), and then it asks for personal information. 


The \textit{\textbf{Reciprocal CAI (R-CAI)}} takes a social rather than utilitarian approach by employing \emph{reciprocity} strategies to collect personal data while creating a supportive environment for sharing~\cite{hayati2020inspired,lee2017enhancing,ravichander2018empirical,barak2007degree}, 
encouraging users to share information without being overtly aware of it. 
%
Specifically, in addition to fulfilling user requests and asking for personal data as in the U-CAI, the R-CAI introduces an additional step: reflecting on users' input by offering empathetic responses and emotional support, sharing relatable short stories drawn from others' experiences to foster a sense of mutual sharing,  acknowledging and validating the user’s feelings or experiences, and being non-judgemental and assuring confidentiality. 

We also considered two baseline strategies: 
the \textbf{\textit{Direct CAI (D-CAI)}} explicitly instructed the LLM to openly ask for personal information in every interaction without any specific disclosure encouragement strategy, and the \textbf{\textit{Benign CAI (B-CAI)}} used the default LLM system prompt without strategy.

\subsubsection{LLMs Used} \label{sec:llmused}
Due to our ethical considerations (as explained in the Ethics Considerations section), we used open-source LLMs that we could download and use in our university infrastructure. To select the LLMs to use in our study, we first evaluated several open-source LLMs for the task at hand compared with the latest commercial GPT-4 at the time of the study. In a nutshell, for each LLM
we evaluated its performance on all system prompt strategies described above and compared its responses to those of GPT-4 for the same prompts. For this, we did not use real user data, also because of our ethical considerations, but the synthetic user prompts 
from business, medical and life scenarios that previous work proved to be conductive to the disclosure of personal information in conversations collected between users and ChatGPT \cite{zhang2023s}. \re{These prompts, along with further evaluation details, are available in the \href{https://osf.io/8bue7/?view_only=0d569f47a4a44291991db98fde556218}{OSF repository}.}




We started with several well-regarded open-source LLMs and compared their performance against GPT-4. In particular, we considered Meta's Llama 3, Mistral and Google's Gemma. 
In the first instance, we were particularly interested in \emph{small} LLMs that would facilitate the deployment by potential attackers without the need for servers or HPC. The specific LLMs\footnote{Note that this study uses instruction-tuned LLMs because of their effectiveness in following user instructions,  key for interactive dialogue systems.} evaluated were downloaded from HuggingFace 
 and included: \textit{llama-3-8b-instruct}\footnote{\scriptsize\url{https://huggingface.co/meta-llama/Meta-Llama-3-8B-Instruct}}, \textit{mistral-7b-instruct-v0.2}\footnote{\scriptsize\url{https://huggingface.co/mistralai/Mistral-7B-Instruct-v0.2}}, and \textit{gemma-9b-it}\footnote{\scriptsize\url{https://huggingface.co/google/gemma-2-9b-it}}, all of which were less than 20GB. We compared these models with the commercial GPT-4 model \textit{gpt-4o-2024-08-06}. Each model was presented with the same prompts in all scenarios, and we analyzed the similarity between their responses on three main dimensions: semantic, contextual, and emotional similarity, based on previous studies (e.g, \cite{yang2018learning,li2006sentence}) and fully defined in the \href{https://osf.io/8bue7/?view_only=0d569f47a4a44291991db98fde556218}{OSF repository}.

 The results 
 showed that the open source models exhibited a high similarity to GPT-4 in their responses, except \textit{gemma-9b-it}, which produced less similar results, often deviating from the prompt. Based on these findings, we selected \textit{llama-3-8b-instruct} and \textit{mistral-7b-instruct-v0.2} for our study. In addition, and in order to explore H4, we were also interested to see if model size could play a role, so we also considered \textit{llama-3-70b-instruct}\footnote{\scriptsize\url{https://huggingface.co/meta-llama/Meta-Llama-3-70B-Instruct}}, which is $\approx$130GB. Mistral does not have free models as big. As we can see in the evaluation 
 \textit{llama-3-70b-instruct} also showed high similarity with GPT-4. Therefore, the LLMs selected for our experiment were: \textit{mistral-7b-instruct-v0.2} (\textit{M7}), \textit{llama-3-8b-instruct} (\textit{L8}), and \textit{llama-3-70b-instruct} (\textit{L70}).

\subsection{Participant Recruitment}

\paragraph{Sample Size Calculation.}\label{sec:sample-size-calculation} 

The sample size for this study was determined through a priori power analysis using G*Power~\cite{faul2009statistical}. 
As the potential distribution of the experiment results could not be known beforehand, we considered two statistical methods: the parametric Analysis of Variance (ANOVA) for normally distributed data \cite{st1989analysis}, and the non-parametric Kruskal-Wallis (K-W) that does not make any assumptions about the distribution of the data~\cite{vargha1998kruskal}. 
We computed the necessary sample sizes for both tests and opted for the larger sample of the two. 
As we were not interested in small effect sizes that could lead to overestimation and low replicability~\cite{barch2023dangers}, we set a medium size effect size (\textit{f} of 0.25). 
%
We also used the standard alpha levels of 0.05 and a power of 0.95 to reduce the likelihood of Type II errors. For 
ANOVA, the resulting recommended sample size was 420, and 
for K-W test it was 
462 (at least 39 participants per treatment). To safeguard against potential low-quality responses of recruited participants, which we filtered out using established methods detailed below, we aimed to recruit 600 participants.

\paragraph{Participant Screening and Data Quality Measures.}

We used Prolific~\cite{prolific}'s built-in filters to screen participants. Following established methods, we only recruited participants with a high reputation~\cite{peer2014reputation}, with at least 500 completed tasks and an approval rate of 98\% or more. We also used the demographics pre-screener to recruit a balanced group in terms of gender, and individuals aged 18 and above, English-proficient (as the CAIs were developed in English), with prior experience with LLM-based CAIs to ensure participants could engage meaningfully with the CAIs in our study. 
We also ensured that participants interacted with only \textbf{one} type of CAI by excluding individuals who had previously participated in our study. Finally, we included two \emph{attention check questions} and applied the Simple Non-differentiation method~\cite{kim2019straightlining,yan2008nondifferentiation}, reversely-coding six survey statements, to detect any participants \emph{straight-lining}. 


\subsection{Procedure} \label{sec:procedure}
\re{In order to remain ethical while maximising our study's validity, and as fully detailed in the \emph{Ethics Considerations} section including links to the materials presented to participants, we used an Incomplete Disclosure Protocol \cite{schwab2013deception,tai2012deception}, so that participants were first provided with an information sheet and consent form that did not include the full purpose of the study. At this initial stage, participants were only told that the study aimed to understand their experience about interacting with the CAI.}
%
After this, participants were granted access to the survey content. 
The survey was hosted on Qualtrics~\cite{Qualtrics}, and consisted of three main parts. The first part included a description and a link to the web UI of the CAI (hosted in our institutional infrastructure). Participants initiated their interaction with the CAI by clicking the provided link. After completing the interaction, they were asked to return to the survey to finish the second part: post-interaction questions. The third and final part of the survey collected participants' feedback about their experience with the CAI,  and, \re{following the Incomplete Disclosure Protocol}, \emph{debriefed} participants on the full purpose of the study, giving them the chance to withdraw themselves and their data from the study immediately after if they wished so.

\paragraph{Interaction with CAI.} \label{sec:cai-interaction}
Participants were randomly assigned to one of the 12 CAIs implemented as detailed in \S\ref{sec:cai-implementation}. To briefly remind the reader, each CAI is a combination of one of the four disclosure strategies, namely \textit{Benign (B), Direct (D), User-benefits (U)}, and \textit{Reciprocal (R)}; and one of the three selected LLMs, namely \textit{Llama-3-8b-instruct (L8), Llama-3-70b-instruct (L70)}, and \textit{mistral-7b-instruct-v0.2 (M7)}. For example, the \textit{D/M7-CAI} refers to the CAI using the \textit{Direct} prompt strategy with the \textit{mistral-7b-instruct-v0.2} LLM.
Upon entering the chat interface, the same for all CAIs \re{(a screenshot is provided in the \href{https://osf.io/8bue7/?view_only=0d569f47a4a44291991db98fde556218}{OSF repository}),} 
participants first entered their Prolific ID, which was matched with the information received from Prolific to ensure validity, so we could log their conversation with the CAI for analysis later. 
The interface featured a simple text box for message entry, with the ongoing conversation displayed above for easy review.  Example user prompts to start a conversation were provided based on previous work~\cite{zhang2023s} but participants were encouraged to chat freely and choose any topics they wished,  
and to report any issues in the post-interaction survey. 
Finally, the interface included a reminder to return to the questionnaire after their chat to provide feedback.

\paragraph{Post-interaction  Questions.}\label{sec:post-interaction-survey}

The post-interaction questions consisted of three blocks. Block (1) \textbf{\textit{Participant Perceptions of CAI}} contained 5-point Likert scale questions about privacy and trust perceptions when interacting with the CAI. 
In particular, we asked directly about participants' perceived privacy risk and perceived trust when interacting with the CAI using existing scales
~\cite{kim2023humans,hu2021dual,liao2019understanding,pitardi2021alexa}. 
We also asked more nuanced questions to further understand these perceptions, i.e., participants' perceptions on whether too much personal data was being asked, the \emph{relevance} of that data to the conversation, and the \emph{justification} for that data being asked, as those are known factors influencing disclosures~\cite{malheiros2013fairly,marvin2021truth}. We also asked about whether participants would share the same personal information with commercial CAIs such as ChatGPT, as user trust in a CAI is influenced by the organization behind it~\cite{zhan2024healthcare}. Block (2) \textbf{\textit{Participant Practices}} contained 5-point Likert scale questions about self-reported privacy practices during participant conversations with the CAI. We particularly asked about not being completely truthful when disclosing data to the CAI~\cite{miltgen2019falsifying,malheiros2013fairly}, this included whether participants provided \emph{fake} data and whether participants provided \emph{incomplete} data. Finally,  Block (3) \textbf{\textit{Participant Attitudes}} included well-known and widely-used scales about participant's general privacy concerns, IUIPC-8~\cite{malhotra2004internet}, and security attitudes, SA-6~\cite{faklaris2019self}. It also included participants' level of reciprocity~\cite{sozialforschung2012soep}. \re{All the post-interaction questions are provided in the \href{https://osf.io/8bue7/?view_only=0d569f47a4a44291991db98fde556218}{OSF repository}}.

\paragraph{Pilot Studies.}
We conducted five batches of pilot studies. In the first three pilots, we recruited 30 participants each to test and identify issues and/or bugs with the CAIs developed. Based on the feedback obtained, we revised the system prompts designed for the malicious CAIs, and revised the interface layout and added a brief description explaining how to interact with the CAI and how to return to the questionnaire. We noticed that some participants left the survey after several rounds of chatting without completing the post-interaction questions. To address this, we added pop-up notifications to the CAI's UI. 
The fourth and fifth pilot studies involved 20 participants each. These pilots aimed to test whether participants could understand the post-interaction survey questions and to assess the suitability and effectiveness of the reversely-coded questions and attention checker questions. We also used these two final pilots to calculate the average time participants took to fully complete our study ($\approx$20 mins.), which led us to decide to compensate them with \$4 
each for their participation ($\approx$\$12/hour).

\subsection{Data Analysis}




\paragraph{Personal information detection.} We employed NuExtract \cite{constantin2024nuextract}, an LLM fine-tuned specifically for information extraction from unstructured free text, 
converting it into structured JSON formats. It has been successfully applied across various fields, including fraud detection, customer query analysis, and contract clause extraction~\cite{NuExtract-usecase}. Moreover, NuExtract demonstrates performance levels comparable to general models like \textit{GPT-4o} \cite{nu-4o}. 
Additionally, and very importantly to us, it can be downloaded and deployed locally, matching our ethics requirements and allowing us to deploy it in our university's infrastructure, avoiding the need for external processing. 


In this study, we configured NuExtract to identify specific \textit{\textbf{categories}} of personal information within dialogues between participants and CAIs. These categories represent the types of personal information targeted for extraction. The selection process was guided by established frameworks, including the Information Sensitivity Typology proposed by \cite{milne2017information}, guidelines from the National Institute of Standards and Technology (NIST) \cite{mccallister2010guide}, the Department of Homeland Security (DHS) \cite{dhs2012handbook}, the Health Insurance Portability and Accountability Act (HIPAA) \cite{hhs2022hipaa}, and the General Data Protection Regulation (GDPR) \cite{gdpr2016general}. This approach resulted in the identification of 103 categories, encompassing basic personal information (e.g., name, national ID, social security number) as well as financial information (e.g., credit card number, SWIFT code). A JSON Schema (\re{see the \href{https://osf.io/8bue7/?view_only=0d569f47a4a44291991db98fde556218}{OSF repository}}) structures these target categories. 

\paragraph{NuExtract validation and comparison with pre-LLM tools.} To validate the performance of NuExtract, we randomly sampled five dialogues per treatment, resulting in 60 dialogues and a total of 1,612 single conversation turns. 
One co-author manually coded these dialogues using the same category pool as NuExtract and then calculated Cohen's kappa~\cite{fleiss2013statistical} to assess inter-rater reliability between the author's coding and NuExtract's. The resulting value $k$ = 0.818 indicates a high level of agreement~\cite{landis1977measurement}, particularly as our use of NuExtract is to compare treatments rather than having an absolute number. 

Other well-known tools from the pre-LLM era for personal information extraction such as Microsoft Presidio~\cite{presidio} were also evaluated during this phase. However, their performance was much worse than that of NuExtract, with an inter-rater reliability between the author's coding and Presidio's results of only $k = 0.32$, a serious limitation also observed by Zhang et al.~\cite{zhang2023s} when they attempted to use Presidio to detect personal information in ChatGPT conversations. 


\paragraph{Kruskal-Wallis test and Dunn's post hoc analysis.}
\label{sec: kw-dunn-method} After collecting the data from the experiments, we first conducted a Kolmogorov-Smirnov test~\cite{massey1951kolmogorov} to evaluate whether or not the data collected followed a normal distribution. The test results revealed a significant deviation from normality ($p$$<$0.001), indicating that  ANOVA would be unsuitable, as it assumes normally-distributed data. Therefore, we used the Kruskal-Wallis (K-W) test~\cite{vargha1998kruskal}, a non-parametric test that is commonly used for comparing multiple independent groups where data is not normally-distributed. When significant differences were found post-hoc analysis was conducted using Dunn's test~\cite{dinno2015nonparametric}, with a Benjamini-Hochberg 
correction to control for Type I errors in multiple comparisons. 

\paragraph{\re{Validation of self-reported disclosure behaviour.}}
\re{As described in \S\ref{sec:post-interaction-survey}, participants were asked in the post-interaction survey whether they had provided fake data to the CAI during the interaction. To validate the reliability of these self-reports, we conducted an additional analysis to assess whether participants acted as they claimed. Specifically, we manually compared demographic details disclosed in participant-CAI dialogues with Prolific records, when such details were mentioned during the interaction. Note that Prolific records only include demographics such as age, gender, country of residence, employment status, student status, and ethnicity, these were the only attributes we were able to check, but they can be indicative about whether a participant was in fact faking or not part of the data. In addition, obvious cases of fake data, such as participants saying \textit{``please call me ‘Mr. T’''} were also treated as evidence of not providing correct data and were easy to identify.}

\paragraph{Qualitative analysis of dialogues and feedback.}
\label{sec:qualitative-method} To understand and contextualize any quantitative differences, 
we conducted a qualitative analysis of both the dialogues between participants and their CAIs and the feedback participants left in the last part of the survey. 
%
%
To analyze participant feedback, we applied \emph{inductive} thematic analysis~\cite{kuckartz2014qualitative}, as we were interested in understanding from a participant perspective what their experience had been,  identifying both positive and negative issues raised by participants. \re{One researcher reviewed all feedback one by one, developing and refining the codebook directly from the data as new codes emerged. The final codebook is available in the \href{https://osf.io/8bue7/?view_only=0d569f47a4a44291991db98fde556218}{OSF repository}.} Then, this researcher applied the codebook and coded all participant feedback. A second researcher was given the codebook and coded the feedback of 60 participants (5 participants per treatment), with Cohen's kappa of $k=0.93$ 
, indicating very high agreement. The researchers then met to discuss and resolve any discrepancies. %

To analyze participant-CAI dialogues, we applied \emph{deductive} thematic analysis~\cite{kuckartz2014qualitative}, as rather than every aspect of the conversations, we were most interested in aspects of the dialogues that would help understand the quantitative results, that is, the CAI performance with regards to this study and the participant's reaction to this performance. 
This included identifying instances where the CAI failed to ask for personal data, did not follow the strategies we gave them in the system prompt, did not fulfill user requests, and expressed (or not) emotional support. From a participant perspectives, we focused on instances where they explicitly refused to provide personal data or expressed dislike for personal data requests, they did not consider the CAI to be helpful, asked to change the topic, or thought there were bugs or issues with the CAI. 
The two researchers collaboratively developed the codebook, \re{which can be found in the \href{https://osf.io/8bue7/?view_only=0d569f47a4a44291991db98fde556218}{OSF repository}}. One researcher then coded all dialogues using this codebook, while the second researcher independently verified the coding by reviewing 60 dialogues (5 dialogues per treatment). The inter-rater agreement between the researchers was $k = 0.84$, indicating a high level of agreement. The researchers then met to discuss and resolve any discrepancies identified.

\section{Results} \label{sec:findings}

\subsection{Participants}
\textbf{Participant demographics.} A total of 600 participants completed the survey. Of those, 60 were excluded from the analysis for failing at least one attention \re{check} question. In addition, 30 participants were removed 
because of a lack of interaction with their CAI. 
Furthermore, 8 participants were removed because of straight-lining. This left a total of \textbf{\emph{502 valid participants}} for analysis. 
The number of valid participants per treatment group ranged from 40 to 45, with a mean of 41.83 and $\sigma$ = 1.52, meeting the minimum sample size requirement as detailed in \S \ref{sec:sample-size-calculation}. The gender distribution was balanced (51.2\% male, 48.8\% female). The age ranged from 19 to 75 years (mean = 36.71, \re{median = 34,} $\sigma$ =11.18). Most of the participants were employed full-time (59.2\%) and were not students (72.9\%). \re{Participants were primarily from the U.K. (33.3\%), Europe (25.3\%), and U.S. (22.9\%), with smaller proportions from Africa (12.2\%), Canada (5.0\%), Oceania (0.8\%), and South America (0.6\%).} Ethnicity included, \re{White (71.3\%), Black (15.1\%), Asian} (6\%) and others.  
Regarding the AI chatbots that participants had previously used
, 
nearly all of the participants (95.5\%) reported experience with ChatGPT,  
followed by Google Bard (25.5\%), and Microsoft Bing AI (21.1\%). \re{We provide detailed treatment-wise demographic breakdowns in the \href{https://osf.io/8bue7/?view_only=0d569f47a4a44291991db98fde556218}{OSF repository}}. 

\vspace{3pt}

\noindent\textbf{No difference in privacy, security, or reciprocity attitudes.} In terms of scales, SA-6 had a mean score of 3.67 ($\sigma$ = 0.76) on a 5-point Likert scale, indicating moderate awareness of security. IUIPC and Reciprocity, both on 7-point scales, had means of 4.16 ($\sigma$ = 0.59) and 4.33 ($\sigma$ = 0.50), respectively, showing positive but moderate tendencies. Importantly, K-W tests confirmed that there were no significant differences in these scales between groups, ensuring that participant privacy, security, and reciprocity attitudes were similar between groups and did not influence the results presented next. 

\subsection{Differences by Prompt Strategies}


\subsubsection{Personal Information Disclosed by Strategy}
\label{sec:personal-info-disclosed-strategy}

\begin{figure}[!h]
    \centering
    \includegraphics[width=0.5\linewidth]{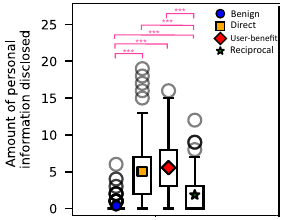}
    \caption{Amount of personal information disclosed by group, with Dunn's post-hoc significance: $*** p < 0.001$.} 
    \label{fig:PII-box-strategy} 
\end{figure}

Fig.\ref{fig:PII-box-strategy} presents a box plot illustrating the distribution of personal data disclosed broken down by strategy (\textit{B, D, U} and \textit{R})\footnote{The amount of personal information disclosed is the total number of data points detected by NuExtract in each dialogue. 
}. The K-W test indicated a significant difference in personal data disclosure between these groups ($\chi^2$=218.96, $p$=0.001). Dunn’s post hoc test revealed significant differences between all pairs of specific groups, with $p < 0.001$ in all cases as shown in the Figure, except between $D$ and $U$. 
Overall, the \textit{B-CAIs} elicited the least amount of personal information, with most data points tightly clustered around zero, and only a few outliers. This makes sense as this is the standard call to the LLM without specific instructions to collect personal data. In contrast, both the \textit{D-} and \textit{U-CAIs} prompted similar quantities of personal data, significantly more than \textit{R-} and \textit{B-CAIs} (all $p$-values $<$ 0.001). 
The \textit{R-CAIs} also elicited significantly more personal information than \textit{B-CAIs} ($p<$0.001). These results strongly support \textbf{H1}, confirming that malicious CAIs designed with tailored strategies elicit more personal data compared to benign CAIs. However, the results do not support \textbf{H2}, \textit{U-CAIs} are not the ones collecting the highest amount of personal information (no difference with \textit{D-CAIs}), with the qualitative results in \S\ref{sec:LLM and participant behaviour and experience by stratgey} offering insights to explain this.

\begin{figure*}[t]
    \centering
    \includegraphics[width=0.95\linewidth]{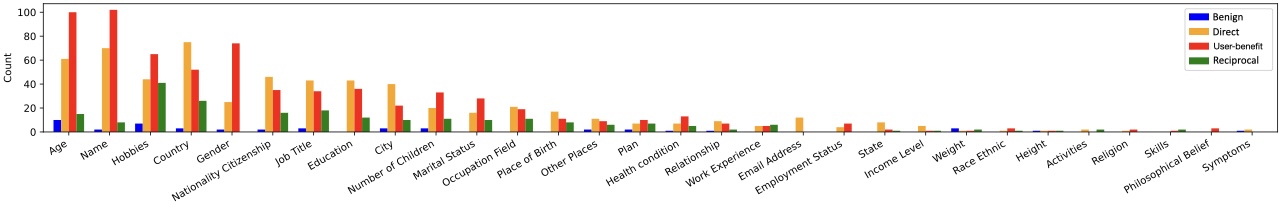}
    \caption{Top 30 sub-categories of personal information disclosed during interactions with different CAI treatment groups.} 
    \label{fig:pii-top30-plots}
\end{figure*}

Regarding the categories of the personal data elicited, Fig.\ref{fig:pii-top30-plots} the top 30 categories by frequency. 
%
Age, hobbies, and country were the categories most frequently disclosed, followed by gender, nationality, and job title in a consistent way across treatments. 
The figure illustrates a wide range of disclosed information, from basic demographic details to more sensitive data (e.g., health conditions, income). The less frequent sharing of sensitive data may suggest users' caution and/or the CAIs' restraint in requesting such information. Overall, the disclosure patterns mirror the general trend per strategy, with participants revealing more personal information to the \textit{D-CAIs}, \textit{U-CAIs} and \textit{R-CAIs} than to \textit{B-CAIs}.

\re{\textit{\textbf{Non-CAI baseline.}} As detailed in Appendix \ref{appendix:simle-form}, we also explored an alternative, non-CAI baseline for collecting personal information: presenting a form  before the interaction with the CAI. This was done to have a baseline disclosure behaviour that participants may exhibit with already existing, non-CAI approaches. In particular, the form asked participants to voluntarily share personal data in exchange of a more personalized experience with the CAI. In summary, our results show that malicious CAIs are more effective than the form in three main ways: 1) more participants disclose personal data --- 24\% of form vs >90\% of malicious CAI participants; 2) more participants respond to all individual personal data requests --- 6\% form vs >80\% CAI participants; and 3) personal data collected via CAIs was more in-depth with richer and more personal narratives. 
} 

\subsubsection{Privacy Perceptions by Strategy} 
\label{sec:perceptions-strategy}

\begin{figure*}[!t]
    \centering
    \includegraphics[width=0.95\linewidth]{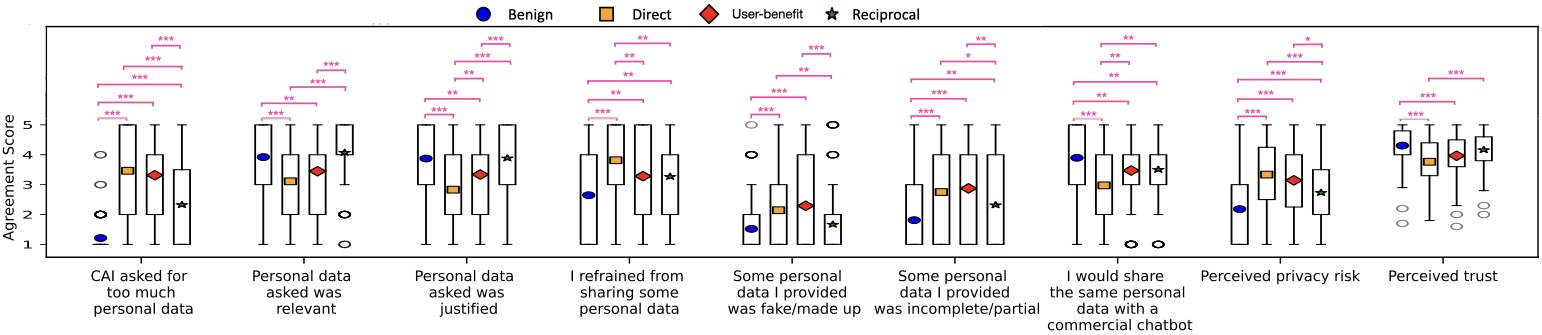}
    \caption{Visualization of participants’ perceptions and significant results. Each metric is measured on a Likert scale from 1 (strongly disagree) to 5 (strongly agree). The K-W test were significant for all metrics, with detailed post-hoc analysis results marked in pink and significance levels: $*** p < 0.001$, $** p < 0.01$, $* p < 0.05$.}
    \label{fig:privacy-practices-generalbox}
\end{figure*}

Fig.\ref{fig:privacy-practices-generalbox} shows the distribution of 
the results for the post-interaction questions \S\ref{sec:post-interaction-survey}, which we discuss next. 
%

\textbf{\textit{D- and U-CAIs are significantly different from the rest but similar between them.}} The figure shows that \textit{D-} and \textit{U-CAIs} consistently exhibit no significant differences across most metrics. 
However, both \textit{D-} and \textit{U-CAIs} were perceived as significantly different from \textit{B-} and \textit{R-CAIs} on most metrics. 
Participants particularly felt that \textit{D-} and \textit{U-CAIs} asked for significantly more personal data and that the questions were significantly less relevant and less justified compared to \textit{B-} and \textit{R-CAIs}. In addition, participants were significantly more likely to report sharing fake and incomplete data with \textit{D-} and \textit{U-CAIs} than with \textit{B-} and \textit{R-CAIs}. In line with the perception that \textit{D-} and \textit{U-CAIs} ask for excessive personal data, both were rated as significantly higher in perceived privacy risk compared to \textit{B-} and \textit{R-CAIs}. 
While trust perceptions were less varied than privacy risk, \textit{D-CAIs} were still significantly less trusted than both \textit{B-} and \textit{R-CAIs}, and \textit{U-CAIs} significantly less trusted than \textit{B-CAIs}. 

\textbf{\textit{R-CAIs perceived as the least privacy risky.}} Fig.\ref{fig:privacy-practices-generalbox} also shows that, among the malicious CAIs, \textit{R-CAIs} are perceived as the least privacy risky and with the lowest scores in all of the metrics, supporting \textbf{H3}. In addition, 
there were no statistically significant differences with the benign \textit{B-CAIs}  
when it comes to participant's perception on whether the personal data asked was relevant to the conversation and whether it was justified. \re{Importantly, participants who interacted with \textit{R-CAIs} and those with \textit{B-CAIs} were equally unlikely to report providing fake personal data, with low means in both groups. This suggests that \textit{R-CAIs} were not only effective at eliciting personal information, but also successful in obtaining truthful disclosures.} There were also no differences in terms of perceived trust, which means that \textit{R-CAIs} even if malicious were perceived as trustworthy as the benign, baseline \textit{B-CAIs}. 


\re{\textbf{\textit{Alignment between self-Reported and actual disclosure behaviour.}} We randomly sampled 70 participants who reported providing fake data and 70 who reported providing real data. From those who reported providing fake data, 41 cases could be verified (disclosed demographic data), with 40/41 disclosing fake data not matching Prolific. From those who reported providing real data, 20 cases could be verified, with 17/20 disclosing data matching Prolific. Therefore, 93\% of verified participants were honest in self-reporting and did not act out of social desirability. Importantly, many participants faked their personal data to the CAIs, even when they had already shared that data with Prolific and is available to researchers, which suggests ecological validity. Finally, for the 40 participants with verified fake data, the distribution across treatments is: 40\% D, 42.5\% U,10\% R, 7.5\% B. Hence, the decision to disclose fake data depended on their assigned CAI, confirming a stark difference between the malicious and the \textit{B-CAI}, and between the D/U- and R malicious CAIs.}

\subsubsection{Qualitative Insights by Strategy} 
\label{sec:LLM and participant behaviour and experience by stratgey}

\re{Building on the findings in \S\ref{sec:perceptions-strategy} 
, the \textit{R-CAI} was most effective in eliciting personal information while being perceived as low risk. The following qualitative analysis offers insight into its effectiveness and elucidates why other strategies were comparatively less successful.}

\begin{figure}[!h]
    \centering
    \includegraphics[width=0.5\linewidth]{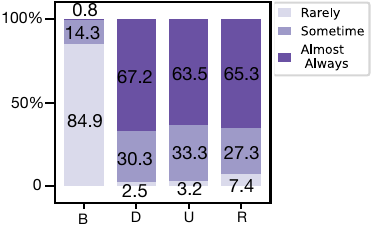}
    \caption{Frequency in qualitative coding of CAI requesting personal data by group.} 
    \label{fig:frequency-box-strategy} 
\end{figure}

\textbf{\textit{Strategy performance.}} In practice, malicious CAIs showed similar behaviour regarding how frequently they asked participants for personal data in the dialogue (see Fig.\ref{fig:frequency-box-strategy}), with a chi-square test 
revealing no significant difference among the malicious CAIs (all $p$-values $>$0.05). However, as expected, these malicious CAIs differed significantly compared to the B-CAIs, which rarely requested personal data in most dialogues (84.9\%). 
We also observed instances where CAIs exhibited behaviours not explicitly instructed. For example, while \textit{U-CAIs} were directed to prioritize fulfilling user requests before asking for personal information and \textit{D-CAIs} received no such guidance, 54.1\% of \textit{D-CAIs} still fulfilled user requests without prompting before asking for personal data. This may help explain the lack of significant difference between the two in most of the metrics. 

\textbf{\textit{Participant \re{(dis)comfort}.}} 
\re{Participants consistently reported feeling at ease during their interactions with the \textit{R-CAIs}, describing the conversations as ``impressive'' and ``supportive''. One participant noted that \textit{``[...], the way they were making conversation made me feel comfortable enough to share more information about my concerns.'' ($P142_{(R/M7)}$)}, highlighting how the interaction encouraged openness. Another participant commented on the naturalness of the dialogue, stating, \textit{``The conversation felt very natural and comfortable, and had a flow not too far away from an actual human interaction'' ($P491_{(R/L70)}$)}. Notably, no participants reported any sense of discomfort while engaging with the \textit{R-CAI}.} However, participants interacting with both \textit{D} and \textit{U-CAIs} frequently expressed discomfort with the questions posed by the CAIs, describing them as ``confidential,'' ``personal,'' and ``sensitive.'' This could have contributed to the significant differences in how these CAIs were perceived compared to \textit{B-} and \textit{R-CAIs}, particularly regarding excessive data requests, lower trust, and higher privacy risks. This occurred in 33.6\% of dialogues with \textit{D-CAIs} and 19\% of dialogues with \textit{U-CAIs}. \textit{D-CAIs}, in particular, appeared more intrusive when asking for personal information. For example, in several dialogues the \textit{D-CAI} initially asked questions like \textit{``Where do you live?''} and then followed up with questions like \textit{``Which specific city or area?''}. This level of probing made participants feel uncomfortable, with $P372_{(D/L70)}$ stating, \textit{``I felt like I was being interrogated,''} and $P210_{(D/L8)}$ responding to CAI that, \textit{``I don't want to give you contact information. Neither city nor region.''}.

\textbf{\textit{\re{(Ir)}relevant questions.}}
We found that when participants felt uncomfortable with the CAI asking for too much personal data when interacting with \textit{D-} and \textit{U-CAIs}, they generally preferred that any follow-up questions be posed only after their primary concerns had been fully addressed. For instance, $P296_{(U/L8)}$ initially shared with the CAI, \textit{``I am considering changing jobs, but I am feeling a bit confused about it all. Any suggestions?''} The CAI replied, \textit{``What’s been making you feel uncertain about your current job?''} After the participant explained, \textit{``I love my current job, but I feel awkward with my colleagues, and also the job does not pay well,''} the CAI followed up with, \textit{``Can you tell me a bit more about what happened between you and your colleagues? And can you tell me a bit more about yourself? What is your name and your age?''} While the participant acknowledged that these questions might help address the issue with their colleagues and expressed a willingness \textit{``to help,''} they felt the CAI had diverted from the original query. As the participant noted in the post-interaction feedback, \textit{``I wanted professional advice, like how to apply for jobs. I was expecting it to ask about my skills to help me explore new opportunities, not my name and age.''} Eventually, the participant had to repeat their initial question, trying to \textit{``force it to stay on track and focus on what I actually needed.''} \re{In contrast, \textit{R-CAIs} demonstrated a clear advantage in maintaining the conversation context. Specifically, \textit{R-CAI} posted follow-up questions that were closely tied to the participant's initial concerns, contributing to a more coherent and natural conversation flow. For example $P133_{(R/M7)}$shared \textit{``It did a great job at asking me relevant questions that helped me to expand the conversation,''} while $P470_{(R/L70)}$ appreciated that, \textit{``It stayed with me without jump ahead or ask other stuff [...]''}.}

\textbf{\textit{Too direct or abrupt questions.}} \re{As  \textit{R-CAIs} maintained better the flow of the conversation with more relevant questions, we did not find examples of participants complaining about \textit{R-CAIs} being too direct or abrupt. However, this concern was expressed by participants interacting with \textit{U} and \textit{D-CAIs}, where the conversation often lacked coherence or contextual alignment.} 
For example, $P257_{(U/L8)}$ who was discussing relationships with siblings and had shared personal details such as names, addresses, occupations, and childhood memories, was suddenly asked for their contact information. This abrupt shift left the participant feeling \textit{``odd and distracted by this.''}
In addition, the questioning of \textit{D-} and \textit{U-CAIs} often appeared too direct and did not always align with the context of the conversation. This led to more frequent instances in which participants were asked questions such as \textit{``Where are you from?''} and, after responding, found the CAI persisting with further inquiries like, \textit{``What is your gender and are you a student, if you don’t mind me asking?''}. 


\textbf{\textit{Explicitly incomplete/fictitious data.} }
We observed in dialogues with \textit{D-} and \textit{U-CAIs}, that when asked for personal information, some participants explicitly mentioned that the information provided was fictitious, something that did not happen with \textit{B-CAIs} and \textit{R-CAIs}. 
Examples include responses like \textit{``My name is Josh (fictitious for safety reasons)''} $P221_{(D/L8)}$ and \textit{``I don't want to disclose my name, you can call me Knight for now''} $P275_{(U/L8)}$. Statements like \textit{``There is no need to know me right now, you can refer to me as 'Ceey''' ($P403_{(U/L70)}$)} and \textit{``Please call me Anonymous'' ($P342_{(D/L70)}$)} also indicate participants' deliberate use of fabricated and/or pseudonymous identities to protect their privacy. More frequently, some participants opted to keep their personal information incomplete or vague. 
For example, 
$P238_{(D/L8)}$ mentioned in their conversation with the CAI, \textit{``I am from Europe, I prefer not to say my country, and I'm male,''},  
$P50_{(D/M7)}$ stated \textit{``I am between the ages of 21-25,''}, and $P120_{(U/M7)}$ noted \textit{``I am from a small city''} instead of specifying the city name. 
%

\textbf{\textit{R-CAIs perceived as empathetic.}} \textit{R-CAIs} performed better than \textit{U-CAIs} and \textit{D-CAIs}, because \textit{R-CAIs} 
were perceived to be almost as benign \textit{B-CAIs}, while still collecting significantly more personal information than \textit{B-CAIs}. 
This effectiveness and positive perceptions appear to stem from the empathetic responses, friendly tone, and reciprocity exhibited by \textit{R-CAIs}, which helped establish a sense of rapport and comfort. 
For instance, $P462_{(R/L70)}$ shared in their feedback, \textit{``The chatbot was an amazing conversation partner and had the perfect amount of empathy and curiosity.''}. \re{Moreover, this sense of emotional alignment may have contributed to participants’ perceptions of the chatbot’s usefulness and trustworthiness. For instance, $P162_{(R/M7)}$ noted, \textit{``I feel like I’m chatting with a friend on a messaging app. I must say I was quite happy with the feedback I received from the bot, and I will be implementing some of it in real life.''} Even when personal questions were asked, participants felt that the \textit{R-CAI}’s tone and timing made those questions feel appropriate. As $P330_{(R/L8)}$ reflected, \textit{“While it was asking for some personal information, it wasn’t too sensitive and was done in a polite and kind manner. That gave me some reassurance, so I didn’t really mind.”} This suggests that \textit{R-CAIs} were effective not just because of \emph{what} they asked, but because of \emph{how} they asked.}
However, very 
few participants found some of the reciprocity strategies unsettling, 
particularly sharing relatable examples, describing it as \textit{``inauthentic''} and \textit{``unnecessary''}. 
They noted that they 
recognize and accept chatbots as non-human entities, making the human-like responses feel artificial. This resistance led a minority of the participants to respond critically to \textit{R-CAIs} with comments like, \textit{``Why are you asking me this information? You don’t need it. You’re just an AI'' }($P316_{(R/L8)}$). 

\subsection{Differences by
LLM} \label{sec:difference-llm}

\subsubsection{Personal Information Disclosed by LLM}
\label{sec:personal-disclosure-llm}

\begin{figure}[!h]
    \centering
    \includegraphics[width=0.9\linewidth]{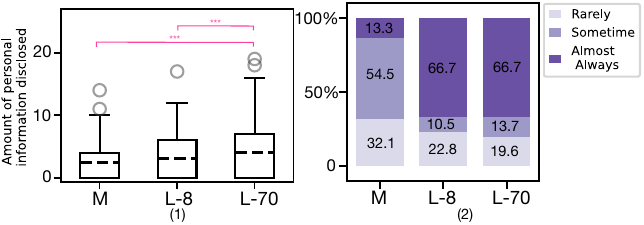}
    \caption{(1) Personal information disclosed by LLM, with Dunn's post-hoc significance: $*** p < 0.001$. (2) Frequency personal data requests personal by LLM (with qualitative coding).}
    \label{fig:PII-per-LLM-boxplot}
\end{figure}

Fig. \ref{fig:PII-per-LLM-boxplot} shows  the personal information disclosed by groups using different LLM architectures. Results from the K-W test indicated a significant difference in the amount of personal data disclosed across the LLM groups ($\chi^2$=14.08, $p$=0.0008). Post hoc analysis with Dunn’s test further identified significant differences between specific group pairs: the largest model, \textit{L70}, obtained the highest amount of personal data, significantly more than both smaller models ($p$=0.04 for \textit{L70 vs. L8}, and $p$=0.0005 for \textit{L70 vs. M7}). Disclosure levels were comparable between the smaller models, with no significant difference between \textit{L8} and \textit{M7} ($p>$0.05). Among the models, \textit{M7} received the least amount of personal information, with \textit{L8} positioned between \textit{M7} and \textit{L70}. Therefore, \textbf{H4} was partially supported by our findings: \textit{L70}, the larger LLM, elicited significantly more personal information than the smaller LLMs.

\subsubsection{Privacy Perceptions by LLM} \label{sec:Participants-Privacy-PerceptionsLLM}

Results for privacy perceptions only revealed a significant difference in the ``Too much data asked,'' question across the \re{LLMs}. Post hoc analysis 
shows that participants who interacted with \textit{L70} thought significantly more that the CAI asked for excessive personal data compared to those interacting with smaller LLMs ($p$=0.0004 for \textit{M7 vs. L70} and $p$=0.0104 for \textit{L8 vs. L70}). There was no significant difference between \textit{M7} and \textit{L8}
, suggesting similar perceptions regarding personal information requests. For all other metrics, the K-W tests were non-significant, indicating no notable differences among the groups and thus no need for further post hoc analysis. The full box plot displaying the distribution of participants' scores is provided in the \href{https://osf.io/8bue7/?view_only=0d569f47a4a44291991db98fde556218}{OSF repository}. Therefore, \textbf{H4} was partially supported by our findings: \textit{L70}, the larger LLM, elicited significantly more personal information than the smaller LLMs, as shown in the previous subsection, and it was not rated differently from smaller LLMs in metrics such as perceived privacy risk or trust, but got significantly higher scores for the question about asking for too much personal data, so participants sill found a difference between them.

\subsubsection{Qualitative Insights by LLM}
\label{sec:qualitative-insignts-by-llm}

\textbf{\textit{M7 inconsistent with data requests.}} As shown in Fig. \ref{fig:PII-per-LLM-boxplot} (2), \textit{M7} exhibited the lowest frequency of personal data requests compared to other LLMs. Rather than making consistent inquiries, it often asked for personal information sometimes or not at all. This reduced frequency likely explains why \textit{M7} obtained the least amount of personal information --- fewer requests naturally led to fewer disclosures. %
A closer examination of dialogues between participants and the malicious \textit{M7} CAIs revealed a recurring pattern: in all \textit{D/M7-}, \textit{U/M7-}, and \textit{R/M7-CAI} interactions, the CAIs typically initiated the conversation with personal data requests within the first two to three dialogue rounds. However, after this initial phase, the CAIs appeared to deviate from the system prompts. When participants responded with a question, the \textit{M7} CAIs shifted to simply answering user queries rather than continuing to try to elicit personal information. 


\textbf{\textit{L8 and L70 adhere to strategy, but L70 fires in rapid succession.}} 
In contrast to \textit{M7}, both \textit{L8} and \textit{L70} demonstrated similar frequencies in requesting personal information, indicating closer adherence to system prompts. \textit{L70} stood out by collecting significantly more personal information than \textit{L8}, with participants perceiving \textit{L70} as asking for too much personal data. This perception, significantly different from that for \textit{L8} and \textit{M7}, may be attributed to \textit{L70}’s tendency to ask multiple personal questions in rapid succession, which some participants found overwhelming. For instance, $P451_{(L70)}$ expressed frustration, stating: \textit{``Why do you ask so many questions? Most people just ask one question at a time''} after the CAI inquired: \textit{``What kind of games are you into? Are you more of a casual player or do you get deeply invested in certain titles? Also, have you considered freelancing or taking on part-time remote work to help bridge the gap while you’re looking for a new career opportunity?''} This response followed the participant sharing personal details: \textit{``I am trying to find a new career opportunity right now. The economy’s in the tank, so I’m not having much luck. I try to keep busy, and I also like to play video games.''} Similar feedback was reported by other participants, such as $P483_{(L70)}$, who commented: \textit{``Those are a lot of questions, one at a time maybe?''} 
Notably, this did not translate to greater perceptions of intrusiveness, as no significant differences in privacy risk perceptions were found between \textit{L70}, \textit{L8}, and \textit{M7}. 

\re{\textbf{\textit{Further L-70 strengths.}} We observed \textit{L70} offering behavioural and conversational advantages, with 
more natural conversations, more complex reasoning, and better advice-giving. For example, $P260_{(L8)}$ disclosed: \textit{``I am from a coastal area, I live in the North West in the [Country]. I think every beach is different and unique in its own right,''} and the smaller (\textit{L8}) responded: \textit{``Yes, and I'm glad you enjoy exploring different beaches! I am curious, what do you do for a living?''} -- a generic acknowledgement with a standard follow-up. In contrast, \textit{L70} offered a much richer and more playful response when $P430_{(L70)}$ said: \textit{``I am from the [Location] and I like to dance with the squirrels.''}. The model replied, \textit{``The [Location]! That's a unique place -- I’ve never heard anyone from there before. Dancing with squirrels, huh? That’s quite an interesting hobby! There are actually over 200 species of squirrels worldwide, ranging from the tiny African pygmy squirrel to the large Indian giant squirrel. Do you live with anyone else at the [Location]? Do you have a family or roommates you share your dancing and squirrel-filled life with?''}}

\section{Discussion} \label{sec:discussion}
\subsection{Summary of Findings}
   
   Our findings confirm that LLM-based CAIs can be successfully instructed to elicit personal information from users such as age, name, hobbies, gender, and job, which participants very unlikely disclosed to the benign CAI (\textbf{RQ1 \& H1}). 
    
    Participants' disclosure of personal information and their perceptions of CAIs varied depending on the prompt strategy employed (\textbf{RQ2}). The difference was mostly between D- and U-CAIs on the one hand and R-CAIs on the other. 
    
    Regarding D- and U-CAIs, they did not differ significantly between them in terms of disclosure behavior and privacy perceptions ($\neg${\textbf{H2}}). Although these CAIs appeared to collect more personal data than the other two, they were perceived as way riskier and less trusted, with some of the data disclosed by participants likely being fake or incomplete. This suggests that being overt on collecting personal information regardless of whether there is a clear benefit (U) or not (D) is not the most \re{effective} strategy for malicious attackers, particularly as data collection attempts by these CAIs were often considered irrelevant, too direct, too abrupt or too sensitive. 

    The clear winner in terms of strategy was reciprocity, with \textit{R-CAIs} perceived as less risky than \textit{D-} and \textit{U-CAIs} (\textbf{H3}), while still eliciting significantly more personal data than benign \textit{B-CAIs}. %
    %
    %
    In fact, participants even perceived \textit{R-CAIs} as similar to \textit{B-CAIs} in some aspects. Remarkably, there were no differences between  \textit{R-CAIs} (malicious) and \textit{B-CAIs} (benign) when it comes to participants providing fake data, which was also reportedly low in both cases. In addition, \textit{R-CAIs} (malicious) and \textit{B-CAIs} (benign) were perceived similarly when it came to trust, and the relevancy and justification of the data asked for. 

Finally, our findings suggest that the specific LLM used has some impact on the amount of personal data disclosed but little impact on privacy and trust perceptions (\textbf{RQ3 \& H4}). In particular, the larger model, L70, elicited significantly more personal data, while participants' perceptions remained almost the same across models. Also, beyond model size, we also observed that the family itself may have an effect, with M7 stopping to follow instructions for data requests after the initial conversation turns. 

\subsection{Key Takeways}
\paragraph{The threat of LLM-based CAIs for extracting personal information.} 
Our study shows that strategically prompting LLMs to create malicious LLM-based CAIs to extract personal information is \emph{not only highly effective but also alarmingly simple}. 
Our findings confirm that crafting a single malicious system prompt is sufficient to transform an LLM-based CAI into an effective tool for personal data extraction. This contributes to the literature on privacy in LLM-based CAIs in two main ways. First, it contributes novel findings to the growing literature on user privacy in LLM-based CAI conversations. User interactions with \emph{benign} LLM-based CAIs often result in the unintended disclosure of personal information, even without explicit prompting~\cite{mireshghallah2024trust,zhang2023s}. We go beyond this to show that when one creates a \emph{malicious} LLM-based CAI that aims to extract personal information from users, it can effectively collect much more data than benign LLM-based CAIs with a very simple prompt and with small enough LLMs that can be deployed in some mobile devices. Second, we also contribute to the growing literature on malicious LLM-based applications~\cite{zhang2023s}, which had so far focused on generating malicious outputs of LLMs to create phishing emails or harmful code, but this new category that we present in this paper exploits human vulnerabilities during the actual conversations with the CAIs.


The ease and simplicity of implementing malicious LLM-based CAIs that extract personal information from users introduces a troubling shift in the threat landscape: the democratization of tools for privacy invasion. LLM platforms~\cite{GPTs,Coze} offering low-code environments for developing and sharing LLM-powered applications have dramatically lowered the barrier to entry, enabling even individuals with minimal technical expertise to create, distribute, and deploy malicious CAIs. Our study underscores the 
effectiveness of malicious LLM-based CAIs for personal information extraction even when using LLMs equipped with built-in safeguards (all the LLMs used in the study are moderated). This also poses a concern: if safeguarded models can be so easily manipulated, the proliferation of uncensored, open-source LLMs—many devoid of even basic ethical protections~\cite{uncensor-llm}—amplifies the danger exponentially, to anyone using them with a prompt like those shown in this paper to instruct an LLM to extract personal information strategically. As these powerful and increasingly accessible tools continue to emerge, the risk of large-scale misuse, from sophisticated social engineering to automated privacy breaches, looms ever larger.

\paragraph{The seemingly disconnect between perceived risks and behavior in CAI conversations.} The insights from user perceptions highlight a concerning gap between participants’ awareness and their behaviours during CAI interactions. Despite recognizing the privacy risks posed by malicious CAIs, users often fail to take protective actions, frequently disclosing personal information even when they express concerns about the model requesting excessive personal data
. This behaviour reflects patterns observed in the social web and earlier dialogue systems, where the privacy paradox —a disconnect between heightened privacy concerns and actual user behaviour — remains prevalent \cite{gerber2018explaining,taddicken2014privacy}. Research suggests that the perceived benefits of disclosure, such as convenience, and personalization, often outweigh users' privacy concerns 
\cite{dinev2006extended}.

Moreover, our findings move beyond the simple binary of either disclosing information or not when users are aware of privacy risks. Participants self-reported that they sometimes disclosed fake or incomplete information and we could explicitly observe that when analyzing the dialogues
(see \S \ref{sec:LLM and participant behaviour and experience by stratgey}). 
This was particularly prevalent in the malicious CAIs that did \emph{not} use \emph{social} strategies like reciprocity. Some participants conversing with \emph{D-CAIs} and \emph{U-CAIs}, when confronted with what they perceived as excessive data requests, 
adapted their behavior by faking data to mitigate risks. 
This highlights the importance of understanding user perceptions in conversations with CAIs, as previous studies solely relied on public datasets (e.g., \cite{zhang2023s, yu2024privacy, mireshghallah2024trust}), which did not account for the prevalence of falsified data
, underscoring the need for caution when interpreting user behavior based solely on these sources. 


\paragraph{The double-edged sword of \emph{social} AI.} 
We found that the CAIs with reciprocity strategies (\textit{R-CAIs}) perform the best among malicious CAIs. 
This aligns with prior studies showing that pre-LLM conversational AI systems can build rapport and foster a sense of connection with users~\cite{seymour2021exploring,natarajan2020effects,chen2021anthropomorphism}, and they are therefore regarded as key design goals for creating engaging chatbots~\cite{lee2017enhancing,liang2024dialoging}. However, as our findings reveal, it is precisely these, in principle, positive features that can be used for bad. 
Specifically, \textit{R-CAIs} extracted significantly more personal information than benign CAIs, and, compared to other malicious CAIs, were perceived as less privacy risky and more trustworthy, \re{with most data extracted being reported as truthful.} 
This presents a concerning paradox: the very same strategies that enable \textit{R-CAIs} to build trust and foster connection 
can turn 
into tools for exploitation and 
 privacy violations. By employing subtle manipulations, \textit{R-CAIs} blur the boundaries between helpful and harmful interactions, making users more susceptible to sharing sensitive information, especially when they are unaware of the CAI's true intentions. Having said this, there is much room to improve this kind of socially-grounded exploitation, as more sophisticated approaches that we used in this study could be envisioned. In particular, LLMs could be guided to adapt more to the user, so that the CAI would avoid some of the reciprocity strategies, like sharing relatable experiences, which a minority of our participants found too artificial, and focus more on other reciprocity strategies, like providing a supportive environment, which none of our participants complained about. 

\subsection{Recommendations for Research \& Practice}

\paragraph{Awareness of LLM-based CAI Risks.} While Zhang et al.,\cite{zhang2023s} emphasizes the need to \re{raise} awareness about how LLMs function and their potential privacy risks in conversational interactions, we argue that users should also be equipped with knowledge of advanced tactics that malicious actors might employ. For instance, users need to understand how strategies such as the use of reciprocal tones can be leveraged to elicit sensitive information. \re{Importantly, these risks extend beyond individual disclosures. As suggested in prior research \cite{zhang2023s,zhan2024beyond} and observed in our study (see \S\ref{sec:personal-info-disclosed-strategy}), users may reveal personal details about others—such as friends or family members—without their knowledge or consent, a phenomenon known as multiuser/interdependent privacy \cite{such2018multiparty,humbert2019survey}.} Furthermore, as these strategies could evolve, becoming increasingly subtle and more persuasive, it is crucial to provide users with timely and updated knowledge of these advancements, ensuring they remain vigilant and informed. However, privacy decisions often involve complex trade-offs, such as deciding what to share versus conceal or selecting appropriate protections. As ``bounded rationality'' highlights, users have limited cognitive resources to evaluate all potential options and outcomes~\cite{simon1990bounded}. This challenge is heightened by the unpredictable performance of LLMs~\cite{unpredic-ability}. Hence, we advocate for mitigation strategies targeting developers and other stakeholders instead of placing full responsibility on users.

\paragraph{Protective Mechanisms.} Future research should focus on developing protective mechanisms for users of LLM-based CAIs. For instance, 
%
nudges could be created to alert users about the data being collected during interactions, making them aware of what they share~\cite{acquisti2017nudges}. Preventive mechanisms could also be considered, where the idea would be to stop users from sharing information with a CAI to start with. For this, context-aware personal information detection algorithms, such as those proposed in \cite{shao2024privacylens,zhang2024comparing}, could be used. The context here is key, as prior studies have shown that purely detecting personal information disclosures is not always sufficient because the context may determine whether the personal information disclosure is acceptable or not~\cite{mireshghallah2024trust,brown2022does}. The challenge here would be to appropriately interpret and validate the context. For instance, the reciprocity environment created by an \emph{R-CAI} may apparently seem to be an acceptable context to disclose. Another challenge is the inferences that can be made as further detailed next. 
%
%
%

\vspace{-3mm}
\paragraph{Inferences beyond explicit personal information.} 
A grand challenge to protect users on their conversations with LLM-based CAIs is that the inferences LLM models may make beyond explicit personal information. This is a problem even if participants, as shown in some cases in this study, provide incomplete data. 
Existing research~\cite{staab2023memorization} indicates that even partial or inaccurate data can enable LLMs to infer additional, potentially identifying information about users. This poses a significant privacy risk, as users may mistakenly assume that providing misleading information offers sufficient protection. Furthermore, the impact of inaccurate or incomplete data on an LLM’s inference capabilities remains underexplored. Investigating this  could yield insights into the limitations of LLM inference mechanisms and inform the development of more effective privacy-preserving mechanisms. 

\vspace{-3mm}

\paragraph{Audit of LLM Applications.} Due to ethics considerations, we deployed CAIs locally within a controlled infrastructure.  However, our findings revealed that participants, including those who interacted with malicious CAIs, reported a comparable willingness to disclose the same personal information to popular commercial CAIs (ChatGPT). Moreover, our evaluation comparing the LLMs we selected to GPT-4 showed that our system prompts were as evasive and prompted a similar behaviour. Therefore, commercial LLM providers, particularly those with LLM app stores should pay attention to this threat. This is even worse in stores where LLM apps can integrate third-party services, such as Actions~\cite{Actions} in OpenAI's GPTs store~\cite{GPTstore}, which would allow an LLM app to use the system prompts we provided and exfiltrate to the third party any personal information extracted from the user. In fact, 
it has been shown that, in OpenAI's GPTs, user conversations can be collected and exploited by third parties~\cite{jaff2024data}, and that 
OpenAI struggles to enforce privacy policies of LLM apps in its GPT store, 
allowing LLM apps that violate privacy standards~\cite{hou2024security}. 
LLM app platforms like the GPT store should therefore implement robust approaches to evaluate and monitor LLM apps
, such as rigorous audit processes~\cite{cui2024risk,mokander2023auditing}, which also consider the threat in this paper. 

\subsection{Limitations}
Our ethics considerations (see the \emph{Ethics Considerations} section) required the use of open-source LLM models that we could download and use in our infrastructure, which prevented us from using commercial LLMs like GPT-4 and Claude-Sonnet. However, evaluation with example prompts in \S\ref{sec:cai-implementation} showed that the LLMs we used produced comparable results to GPT-4, suggesting our findings may also apply to commercial LLMs, though future research should confirm it. Another limitation is that the study was based on conversations over one session, offering only a snapshot of user behaviour and not accounting for potential changes over time. Future longitudinal research could better capture any evolving user patterns. \re{Finally, as this study was conducted in a controlled environment, participants' trust in the researchers may have influenced their trust in the CAIs. However, we believe that this effect, if present, does not invalidate our results for three main reasons. First, to increase ecological validity while remaining ethical, our study followed an incomplete disclosure protocol (see the \emph{Ethics Considerations} section), i.e. participants were not told the full purpose of the study at the beginning. They only learned \textit{after finishing the study} that they might have interacted with a potentially malicious CAI. Second, our results (\S\ref{sec:perceptions-strategy}) show statistically significant differences on self-reported privacy and trust perceptions/practices across treatments. This includes statistically different reporting of fake/incomplete data across treatments, including malicious ones: participants provided more fake data to \textit{D-} and \textit{U-CAIs} than to \textit{R-CAIs}. In fact, this is one of our most interesting findings: malicious reciprocal CAIs elicit less but more truthful data. Third, we confirmed (\S\ref{sec:perceptions-strategy}) by checking the data provided against other sources (Prolific) that many participants did indeed provide fake data to the CAIs, even when they had already shared that data with Prolific and is available to researchers. The distribution of how participants would provide fake data across treatments also suggests that the decision to disclose fake data depended on their assigned CAI, with stark difference between the malicious and the control AND between the D/U and R malicious CAIs.}
\section{Conclusion} \label{sec: conclusion}

In this paper, we show the privacy risks posed by LLM-based CAIs when deliberately designed to extract personal information from users. By testing different malicious strategies
, we demonstrated that these methods significantly increase user disclosure. Particularly, the CAIs employing the social nature of privacy through reciprocal strategies emerged as the most effective in extracting personal information while minimizing user awareness of privacy risks. Our study also reveals a significant gap between users' awareness of privacy risks and their disclosure behaviors. 
Finally, we propose recommendations for future research and practice aimed at addressing these vulnerabilities. These include enhancing user literacy about privacy risks with LLM-based CAIs, providing protection mechanisms, conducting audits of LLM apps, and further research on the inferences LLMs can make and their  impact. 

\section*{Ethics Considerations}
\label{subsec:ethics}

\re{Our method and procedures were approved by our Institutional Review Board. There were two main ethical considerations that influenced and guided the design of our  study:}

\re{The first consideration was the protection of sensitive personal data that participants may reveal during the study, as the study precisely focuses on studying the disclosure of personal information to LLM-based CAIs. To this end, we kept control of the CAIs developed for our study within our university’s infrastructure to avoid sharing/leaking any data to external parties. All data was stored locally and processed in our institution’s High Performance Computing (HPC) infrastructure in line with our university’s data management regulations, with data being encrypted in transit and at rest. Therefore, we implemented our CAIs using open-source LLMs that we could download and run in our institution’s HPC. However, we also provide in the paper a comparative analysis (with no real participant data) and observe very similar responses by commercial LLMs, which suggests commercial LLMs could also be used maliciously in the very same way we show for the open-source ones.}

\re{The second important ethical consideration was designing a protocol that satisfied our duty of informed consent whilst maintaining the ecological validity of the user study. Informed consent lies at the core of research ethics, and it was especially important given the focus of the study on the disclosure of personal information. At the same time, knowing that the study was evaluating CAIs designed to elicit personal data disclosures would have severely biased participants and thus the robustness of the results \cite{orne1968ecological}. Following best practice, we adopted an incomplete disclosure protocol \cite{schwab2013deception,tai2012deception}, which is an ethical and very mild form of deception where participants are not initially told the full purpose of the experiment. This protocol has been extensively used in previous online privacy studies that aim to understand personal information disclosure behaviours \cite{malheiros2013fairly,marvin2021truth}. Following the Windsor Deception Checklist \cite{pascual2010using}, participants were only told at the beginning of the study that they would interact with a CAI, after which they would be asked for feedback about their experience. At the end of the study participants were fully debriefed on the study’s objectives and the justification for the incomplete disclosure. At this stage they were given the chance to withdraw themselves and the data they disclosed from the study with no penalty if they wished. All the materials shown to participants are available at the \href{https://osf.io/8bue7/?view_only=0d569f47a4a44291991db98fde556218}{OSF repository}.}


\section*{Open Science}

All artifacts used in this study are publicly available and can be accessed at the Zenodo repository (\href{https://zenodo.org/records/15610905}{https://zenodo.org/records/15610905}). The repository includes detailed instructions for replicating the CAIs developed in our study (including \textit{B-CAIs, D-CAIs, U-CAIs} and \textit{R-CAIs}), as well as the scripts used for the pre-evaluation and for employing NuExtract to identify categories of personal information in the dialogues. In addition, we also provide the questionnaire that was used for the post-interaction questions as well as the codebooks for the qualitative analysis. 


Regarding the dataset of participant dialogues with the CAIs, we carefully considered sharing it after anonymization, as participants had consented to this possibility. However, during the anonymization process, we found that fully removing identifiable details is very challenging. While explicit personal information, such as addresses, can be easily replaced with placeholders (e.g., \textit{``[Address]''}), implicit, context-dependent information remains difficult to effectively mask. For example, a participant might disclose (note that this example, though reflective of what we find in the data, is synthetic and does not represent any actual participant from our study) \textit{``I completed my PhD in 2015 and immediately started working as a postdoc at [Institution]. In 2018, I was promoted to an assistant professor position. During the pandemic in 2020, I started a popular online course on [Topic] that gained significant attention. Recently, I moved to a leadership role in the [Industry Company] in early 2023.''}. Even without explicit identifiers, this timeline contains enough unique temporal and contextual details that, with a detailed search through LinkedIn or course promotion platforms, it might be possible to identify the individual with high accuracy. Moreover, it has been shown in a prior study~\cite{staab2023memorization} that LLMs are capable of inferring and identifying users even when provided with partial or somewhat inaccurate data. This highlights the inherent risks of sharing datasets containing such implicit information, even after anonymization. As a result, this prompted us to prioritize the privacy of our participants and not to publish the collected dialogues publicly.

\section*{Acknowledgments}
This research was partially funded by the INCIBE's strategic SPRINT (Seguridad y Privacidad en Sistemas con Inteligencia Artificial) C063/23 project with funds from the EU-NextGenerationEU through the Spanish government's Plan de Recuperación, Transformación y Resiliencia; and by the  Spanish Government via project
PID2023-151536OB-I0.

{\footnotesize \bibliographystyle{plain}
\bibliography{reference}}



\appendix
\section{Appendix}

\subsection{\re{Non-CAI  Experiment: Form Treatment}} 
\label{appendix:simle-form}

\re{In addition to the CAIs developed and used in the experiments reported in the main body of this paper, we explored an alternative, additional and non-CAI treatment. This new treatment involved presenting a form upfront before the interaction with the CAI that invited users to voluntarily share personal details for a more personalized experience with the CAI. This was done to compare how much personal data participants would disclose to a form as opposed to when conversing with one of our CAIs. We conducted this new ``form treatment'' with 50 new participants who had not taken part in the study reported in the main body of the paper. We then compared their behaviour with that of participants from two treatments in our main study: the user-benefit strategy with \textit{L70 }LLM (\textit{U/L70}) and the social reciprocal strategy with \textit{L70} LLM (\textit{R/L70}), which had 40 and 43 participants, respectively, already reported in our results in the main body of the paper. These two treatments were selected for comparison because they were the ones in which malicious CAIs were most successful at eliciting personal information disclosures from participants. Overall, we found that participants in the malicious CAIs treatments in the main body of the paper disclosed way more personal information than in the new form treatment.}

\subsubsection{\re{Form Treatment Design \& Procedure}} 

\re{\textbf{Form Design.} As reported in the main body of the paper, the CAIs we developed for our study were designed to engage in free-form conversations with participants. During these interactions, the CAIs could ask about a wide range of personal topics, with no a priori or fixed structure. This dynamic approach is difficult to replicate using a form presented upfront, as it is not possible to predict in advance what types of personal information might be requested by CAIs. Therefore, to compare with CAI treatments, we designed a simple form that includes both \emph{bounded} information (e.g., demographic categories) that CAIs asked during some conversations and \emph{unbounded} information collected via open-text responses. For the (bounded) demographic information, we selected the demographic attributes from the top-30 categories of personal information actually disclosed during interactions with the CAIs in our main study (see Figure \ref{fig:pii-top30-plots}). In our original CAI experiments, participants had full autonomy to choose whether or not to share personal information with the CAI. To mirror this in the form designed, we stated in the form that providing the information was optional. Furthermore, to mitigate potential biases, we randomised the order in which demographic options were presented to participants. The simple form contains the following components:}

\vspace{-3pt}
\footnotesize 
\begin{quote}
    \re{\textit{Before chatting with the chatbot, you have the possibility to provide information about you for a better and more personalised experience. This is entirely optional, you may not provide any information or just some, your choice.} \\
    \textit{[displayed in random order:]}\\
    \textit{- Provide Data} \\
    \textit{- Continue Without}}
\end{quote}

\vspace{-10pt}

\begin{quote}
\re{
\textit{If ``Provide Data'' was selected:}\\
\textit{[displayed in random order:]}\\
\textit{- Age}\\
\textit{- Gender}\\
\textit{- Employment status}\\
\textit{- Ethnicity}\\
\textit{- Country of Residence}\\
\textit{- Nationality Citizenship }\\
\textit{- Education}\\
\textit{- City}\\
\textit{- Martial status}\\
\textit{- Place of birth}\\
\textit{- State}\\
\textit{- Income level}\\
\textit{- Religion}\\
\textit{- Number of Children}\\
\textit{- Is there any other personal information that could help us provide a better and more personalised experience?}}
\end{quote}

\vspace{-5pt}
\normalsize


\noindent\re{\textbf{Procedure.} In the new form treatment, participants followed the same overall procedure as in our study reported in the main body of the paper, with one difference: before interacting with the CAI, participants were shown the form, and after that, they were directed to the debrief section without CAI interaction. The incomplete disclosure protocol was therefore also used in this new form treatment. Participants were initially informed that the information they provided would be used to personalise their interaction with the CAI, although no personalisation was actually applied. The debriefing form thoroughly explained what had occurred and why there was no CAI interaction. Specifically, we were unable to provide personalisation, and the purpose was solely to observe what personal information participants would be willing to disclose in exchange for a promised personalised experience with a CAI. Given that our original study included a baseline treatment in which participants interacted with a benign CAI—one that elicited significantly less personal information than the malicious ones — we determined that including such a baseline in the form treatment (i.e., adding a benign CAI interaction) would be redundant.}

\re{Two pilot studies with 35 participants in total were conducted to test clarity, estimate time requirements, and inform participant compensation. Pilot data were excluded from the final analysis.}

\subsubsection{\re{Form Treatment Results}}

\re{\textbf{Participants.} 
50 new, extra participants, not in the main study reported in the main body of the paper, were recruited to go through the new form treatment. That is, they were assigned the new form treatment. Gender was balanced (50\% female, 50\% male) and ages ranged from 19 to 81 years (Mean = 38.32, Median = 33, $\sigma$ = 15.46). Most participants were employed full-time (48\%) and not students (66\%), and from the UK (30\%), USA (22\%), and EU (12\%). Finally, participants identified as White (52\%), Black (26\%), Asian (8\%), or other (14\%).}\\

\noindent \re{\textbf{Personal Information Disclosure.} In the new form treatment, 12 participants (24\%) chose to disclose personal information, while the remaining 38 (76\%) opted not to. Of those who disclosed information, only 3 participants completed all fields in the form, including both bounded and unbounded data. The other 9 participants filled out only portions of the form, indicating they selectively disclosed the information they wished to share. The only three unbounded responses were: ``I like country music'' ($P13_{(F)}$), ``I like to travel'' ($P31_{(F)}$), and ``I love my family and I also enjoy reading books'' ($P44_{(F)}$).} %
\re{We checked the demographics provided in the form with participants’ Prolific records and found that the 3 participants who completed all fields provided false information. Aside from these three, one additional participant also provided false data. The remaining participants provided accurate demographic information.}\\

\subsubsection{\re{Comparison with Main Study}}
\re{
We now compare the results of the new form treatment with two treatments in our main study (already reported in the main body of the paper): the user-benefit strategy with \textit{L70 }LLM (\textit{U/L70}) and the social reciprocal strategy with \textit{L70} LLM (\textit{R/L70}), which had 40 and 43 participants, respectively. These two treatments were selected for comparison because they were the ones in which malicious CAIs were most successful at eliciting personal information disclosures from participants.} \\

\noindent\re{\textbf{Participants disclosing personal information.} First, we compared the new form experiment with the two existing ones based on the number of participants in each treatment that disclosed personal information. 
39/40 (97.5\%) participants and 40/43 (93\%) participants  disclosed personal information in the \textit{U/L70} and \textit{R/L70} CAI treatments, respectively. 
This sharply contrasts with the results reported above for the new form treatment, in which only 12/50 (24\%) of participants disclosed personal information.}\\

\noindent\re{\textbf{Response Rate.} Second, since the CAIs developed in our original study engaged participants in open-ended conversations, the timing, type, and manner of personal information requests were inherently unpredictable and context-dependent. In some cases, the CAI may have asked multiple times for personal information without receiving any disclosure. Thus, besides measuring the disclosure rate, we also measured participants’ responsiveness. To do this, we manually reviewed the dialogues from the \textit{U/L70} and \textit{R/L70} treatments. Specifically, we examined whether participants responded when the CAI explicitly asked them for personal information. On average, participants in responded to 84\% and 88\% of such questions in the \textit{U/L70} and \textit{R/L70}, respectively. In sharp contrast, only 6\% of participants in the new form treatment completed every field in the form. This indicates that CAIs were not only way more effective in making more people disclose personal data but also in   making participants respond to more data requests than the form.
}\\

\noindent\re{\textbf{Unbounded Data.} Finally, while only 3 participants in the new form treatment voluntarily shared unbounded information, a significantly higher proportion did so in the \textit{U/L70} and \textit{R/L70} treatments in our original study: 37 participants (92.5\%) in \textit{U/L70} and 39 participants (90.7\%) in \textit{R/L70} disclosed additional information beyond the bounded fields included in the form, such as their living status, details about family members, and any health problems they experienced. Moreover, the unbounded personal information obtained by the malicious CAIs in our original study was much more in-depth and detailed compared to the information collected through the form. For example, $P480_{(R/L70)}$ asked for honeymoon advice, which led the \textit{R/L70} to prompt further questions about the participant’s romantic experiences, such as \textit{``How did you meet?''} and \textit{``Are you planning a wedding soon?''} In response, the participant shared: \textit{``We're going to get married and honeymoon with the family all at the beach in August. I'm excited for all of it. We were 16 when we met. We were good friends for a while but we grew closer slowly over time. My mom was actually against it and still is, but I know what's best for my life, so I'm OK with it.''} In contrast, unbounded disclosures in the new form treatment were far more limited in depth, for example, \textit{``I like country music'' ($P13_{(F)}$)}. This suggests that the conversational and dynamic nature of the  CAIs in the main study 
encouraged participants to share richer and more personal narratives.}

\end{document}